\documentclass[twocolumn]{aastex63}
\usepackage{amsmath}
\usepackage{amssymb}
\usepackage{float}
\usepackage{graphicx}
\usepackage{multirow}
\usepackage{booktabs}  
\usepackage{threeparttable}
\usepackage{enumitem}
\usepackage{lineno}

\newcommand{\BH}[1]{\textcolor{black}{#1}}
\newcommand{\ZL}[1]{\textcolor{black}{#1}}

\newcommand{\lya}{Ly$\alpha$}
\newcommand{\dr}{$\delta_F^{\rm{rec}}$}

\received{Feb 24, 2021}
\revised{April 14, 2021}
\accepted{May 11, 2021}
\submitjournal{ApJ}

\shorttitle{ORCA: Optimized Reconstruction with Constraints on Absorption}
\shortauthors{Li, Horowitz, and Cai}
\graphicspath{{./}{figures/}}

\begin{document}

\title{Improved Lyman Alpha Tomography using Optimized Reconstruction with Constraints on Absorption (ORCA)}


\correspondingauthor{Zheng Cai}
\email{zcai@mail.tsinghua.edu.cn}

\author{Zihao Li}
\affiliation{Department of Astronomy, Tsinghua University, Beijing 100084, People's Republic of China
}
\affiliation{Department of Aerospace Engineering, Sichuan University, Chengdu 610207, People's Republic of China}


\author{Benjamin Horowitz}
\affiliation{Department of Astrophysical Sciences, Princeton University, Princeton, NJ 08540, USA}

\author{Zheng Cai}
\affiliation{Department of Astronomy, Tsinghua University, Beijing 100084, People's Republic of China
}

\begin{abstract}

In this work, we propose an improved approach to reconstruct the three-dimensional intergalactic medium from observed Lyman-$\alpha$ forest absorption features. We present our new method, the Optimized Reconstruction with Constraints on Absorption (ORCA), which outperforms the current baseline Wiener Filter (WF) when tested on mock Lyman Alpha forest data generated from hydrodynamical simulations. We find that both reconstructed flux errors and cosmic web classification improve substantially with ORCA, equivalent to 30-40\% additional sight-lines with the standard WF. We use this method to identify and classify extremal objects, i.e. voids and (proto)-clusters, and find improved reconstruction across all summary statistics explored. We apply ORCA to existing Lyman Alpha forest data from the COSMOS Lyman Alpha Mapping and Tomography Observations (CLAMATO) Survey and compare it to the WF reconstruction.

\end{abstract}


\section{Introduction} 
\label{sec:intro}
The galactic superclusters, sheets, filaments and voids are a significant part of our observable Universe, which are also referred to as the large scale structure of the universe. Cosmography is the science of mapping and describing those features using a variety of probes. While the local universe can be mapped out with galaxy surveys like SDSS\citep{Eisenstein_2011}, all sky galaxy surveys are restricted to the universe at low redshift since the surface brightness scales with redshift as $\propto(1+z)^{-4}$, and we even no longer resolve a galaxy disk at longer distances, making it increasingly difficult to map the large scale structure at higher redshift.

An alternative probe of large scale structure at $z>2$ is Lyman-$\alpha$  forest absorption in spectra of background quasars and galaxies, which is complementary to low redshift galaxy surveys. The Ly$\alpha$ forest traces the neutral hydrogen density \citep{1965ApJ...142.1633G} which can correspond to the underlying dark matter density through fluctuating Gunn-Peterson approximation (FGPA) \citep{1998ApJ...495...44C}. While the flux along a single line-of-sight (LOSs) towards a background source (quasar or galaxy) only provides one dimensional information, 3-D map can be reconstructed using inversion methods given a set of LOSs toward a group of background objects \citep{2008MNRAS.386..211C}. A long-used method for this purpose is Wiener Filter \citep{2001MNRAS.326..597P}, which has been validated to recover the dark matter field by \citet{2008MNRAS.386..211C} and the observational requirements for implementing such method is discussed by \citet{Lee_2014}. Since this early work, the Wiener Filter has been widely used in many surveys including CLAMATO \citep{Lee_2018},  and LATIS \citep{Newman_2020}, \ZL{and eBOSS-Strip 82 \citep{Ravoux_2020}, the largest \lya\ tomography map ever made so far.} Going forward, there is significant interest in expanding \lya\ tomography as a probe of the IGM over more sky area and cross-correlate the properties of the IGM with overlapping galaxy samples, such as in the already planned Subaru Prime Focus Spectrograph (PFS) \citep{2014PASJ...66R...1T} galaxy evolution component and the more futuristic proposed surveys on facilities like TMT/ELT/GMT.

In addition to Wiener Filter, there has been recent interest in reconstructing the underlying matter density field associated with the observed flux through a foreword modeling framework \citep{Horowitz_2019,horowitz2020tardis,2019A&A...630A.151P}. Unlike Wiener Filter, TARDIS \citep{Horowitz_2019} reconstructs the initial density field through an optimization framework powered by a fast differentiable particle-mesh solver (either FlowPM \citep{modi2020flowpm} or FastPM \citep{2019ascl.soft05010F}) and convert the evolved matter density field to a flux field assuming the analytical Fluctuating Gunn-Peterson Approximation (FGPA). These methods get the final reconstruction by finding the maximum \textit{a posteriori} initial density field which gives rise to the observed field. A unique advantage of TARDIS is that, owing to gravitational evolution, it yields both information on the underlying dark matter density field and on velocity allowing us to deconvolve redshift space and real space quantities and provide a more accurate reconstruction. However, such a method depends on N-body simulation and FGPA model which inherently rely on cosmological and astrophysical assumptions. These methods reconstructed fields all within known frames and this may omit some unknowing physics (e.g., the underlying true universe may not be well modeled by the current cosmological model we use).

One important feature of large scale structure is void, which occupies the majority of the volume of cosmic web. Voids are regions lacking large galaxy populations and their density are below mean cosmic density \citep{doi:10.1142/S2010194511000092}. However, understanding and cataloging these regions have been found to be important for a number of cosmological analyses including constraining neutrino properties \citep{2019MNRAS.488.4413K,2020PhRvD.102h3537Z, 2020PhRvD.101f3515L}, dark energy \citep{lee2009constraining, bos2012darkness}, and modified gravity \citep{2019A&A...632A..52P,2020arXiv200903309C}. Cosmic voids catalogs have been constructed from spectroscopic surveys \citep{mao2017cosmic}, \ZL{tomographic maps \citep{Krolewski_2018}} and photometric surveys \citep{2019MNRAS.490.3573F}. However, identifying cosmic voids at high redshift with statistical significance is difficult using galaxies themselves due to the low number density and high sample variance. 

The counterpart of voids are galactic (proto)-clusters, which are important for studying star formation history and the origin of large scale structure. Observations of low redshift galaxies in cluster environments have shown older stellar population and lower star formation rates than those located in the field \citep{2005Wake,2009Skibba}, indicating that these cluster environments underwent a cycle of significant star formation and quenching at high redshift ($z>2.0$) \citep{2010Tran}, so-called ``Cosmic Dawn." 
Deep galaxy redshift surveys provide a promising way to study these (proto)-clusters, for example in the COSMOS field \citep{Chiang_2014}.
However, similar to void discovery, identifying (proto)-clusters at high redshift is difficult using galaxies resulting in finding only the most massive protoclusters in the deepest fields (such as COSMOS). Observations with Ly$\alpha$ forest provide a useful alternative for identification of proto-cluster environments, either through tomographic reconstruction 
(e.g.\citet{2015MNRAS.453..311S}) or through a group of spectra analysis  within a protocluster scale \citep{2016ApJ...833..135C,Cai_2017}.




The properties of cosmic voids and proto-clusters are the result of non-linear structure formation from the Gaussian early universe to the time of observations. However, the Wiener Filter only provides an unbiased minimum-variance estimate in the limit that the underlying field is Gaussian \citep{tegmark1997measure}. While this is a valid approximation for power-spectra analysis of the CMB \citep{1998PhRvD..57.2117B} or galaxy surveys \citep{vogeley1996eigenmode,1998ApJ...499..555T}, when studying the structure of non-linear features one expects there to be additional information that is not captured or is otherwise smoothed by the filtering. With next generation facilities coming online later this decade, \lya\ tomography will be possible across large regions of the sky at comparatively spatial resolution ($\sim 1$ Mpc/h), probing deeper into this nonlinear regime.

We review the Wiener Filter and introduce our extension, Optimized Reconstruction with Constraints on Absorption (ORCA), in section \ref{sec:reconstruction}. We apply both Wiener Filter and ORCA to mock survey with Nyx simulation in section \ref{sec:mock surveys}, and we compare the accuracy of cosmic web classification in \ref{sec:cosmic web} to show the improvement of ORCA. In section \ref{sec:clamato data}, we apply ORCA to CLAMATO data and compare our new results to previous works done with Wiener Filter. We further apply the spherical void and cluster finder algorithm in section \ref{sec:vc finding} and we present the void and cluster profiles in different maps, finding reconstruction of ORCA more consistent with Nyx simulation. In section \ref{sec:conclusion}, we discuss the results of this paper.
In this paper, we assume a flat $\Lambda$CDM cosmology, with $\Omega_M=0.31,\Omega_\Lambda=0.69$ and $H_0=70\ {\rm km}\ s^{-1}\ {\rm Mpc}^{-1}$.



\section{Reconstruction methods} \label{sec:reconstruction}

\subsection{Wiener Filter}
The Wiener Filter is a special case of the maximum likelihood method, which analytically reconstructs the field with maximum likelihood given observed data with known priors (e.g., the underlying distribution of the data). 
In the standard Wiener Filter approach, the reconstructed map 
is estimated from the data d by evaluating
\begin{equation}
M=C_{MD}\cdot(C_{DD}+N)^{-1}\cdot d,
\end{equation}
where $C_{MD}$ and $C_{DD}$ are the map-data and data-data covariance matrix, $N$ is the noise covariance matrix, which is diagonal assuming noise is uncorrelated, and $d$ is the input data. The shape of covariance matrix can be calculated in different approaches with cosmological priors, and in a widely accepted \emph{ad hoc} approach, a Gaussian random linear filed prior is assumed \citep{2001MNRAS.326..597P}. As is implemented in \citet{2008MNRAS.386..211C,Lee_2014}, covariance of two points  $\boldsymbol{r_1}$ and $\boldsymbol{r_2}$ either in map and data is Gaussian, so that $C_{MD}$ and $C_{DD}$ can have the same definition: $C_{MD}=C_{DD}=C(\boldsymbol{r_1},\boldsymbol{r_2})$ and
\begin{equation}
C(\boldsymbol{r_1},\boldsymbol{r_2})=\sigma^2_F exp\left[-\frac{(\Delta r_{\parallel})^2}{2L^2_{\parallel}}\right]exp \left[-\frac{(\Delta r_{\bot})^2}{2L^2_{\bot}}\right]\label{con:wiener}
\end{equation}

where $\Delta r_{\parallel}$ and $\Delta r_{\bot}$ are the distance between $\boldsymbol{r_1}$ and $\boldsymbol{r_2}$ along, and transverse, to the line-of-sight, respectively, and $\sigma_F$ is the the priori expected variance of  3-D Ly$\alpha$ Forest flux fluctuations in a volume of order $L^2_{\bot}L_\parallel$, while $L_{\parallel}$ and $L_{\bot}$ are correlation lengths along and perpendicular to the LOSs. $L_{\bot}$ is often chose to be on the order of LOSs mean separation $\left\langle d_{\rm{LOS}} \right \rangle$ to avoid fictitious structures while $L_{\parallel}$ depends on the specific scenario (e.g., \citet{Lee_2014} set $L_{\parallel}$ to FWHM of the assumed instrumental
resolution while \citet{2008MNRAS.386..211C} set it to be of the order of the Jeans length in order to avoid information loss for small scales along the LOSs). The choice of all parameters we use in this paper are discussed in Section \ref{sec:flux recon}. We use the Wiener Filter codes {\tt dachshund}\footnote{\url{ http://github.com/caseywstark/dachshund}} developed by \citet{2015MNRAS.453..311S}.

In practice,  $d$ is a column vector containing observed flux contrast from all lines of sight, $\delta_F=F/\langle F \rangle-1$, while $(C_{DD}+N)$ and $C_{MD}$ are two large matrices containing correlation information. Although the noise matrix $N$ is usually assumed to be diagonal, the covariance matrix $C_{DD}$ is complicated as the signals are correlated to each other, which makes it tremendously computational intensive to inverse $(C_{DD}+N)$ as our surveys become larger. \ZL{\citet{2015MNRAS.453..311S} has implemented an iterative pre-conditioned conjugate gradient (PCG) method in {\tt dachshund}, which reduces the time complexity to $O(N^2)$ and space complexity to $O(N)$}. There are also attempts to reduce computation like dividing the box into small overlapping chunks \citep{Lee_2014}, but there are still significant costs associated with the matrix to inverse operation.

Note that Wiener filtering only matches the maximum a posteriori estimator in the case of a Gaussian random field whose properties are described solely by the signal covariance \BH{so we shouldn't expect it to be optimal when those conditions are not met}. Even without taking into account the nonlinear evolution of the matter density of the universe, this will not describe the flux field at $z \sim 2.0$ as hydrodynamical effects are nonlinear in the underlying density field. This is particularly true in (proto)cluster regions where feedback effects are expected to be strong. \BH{Since the Gaussian approximation is not valid for the underlying flux field, we expect it should be possible to construct an alternate estimator which outperforms the Wiener filter estimate across different cosmic environments.}

\subsection{ORCA}
As standard Wiener Filter requires the inversion of a 
large matrix, we can approach the reconstruction as 
an optimization problem using an L-BFGS optimizer to 
avoid intensive computation, which is also used 
in \citet{Horowitz_2019}. While \citet{Horowitz_2019} 
reconstruct the initial density field using 
Gunn Peterson Approximation, we are directly reconstructing 
the flux field without such an assumption, allowing 
generalization to different cosmological models without 
residual biases.

An estimate of the underlying flux field\footnote{\ZL{We refer flux to transmitted flux in this paper. The transmitted flux or transmission is the observed flux divided by continuum: $F=f/C$.}}, $s$, can be found by minimizing the loss function $\mathcal{L}$,
\begin{equation}
\begin{split}
\mathcal{L} &= k_1(S_m(s) - s)^2\\
&+(R(s) - d)^\mathrm{T}N^{-1} (R(s)-d)\\
&+k_2\sum clip(s,1,+\infty)+k_3\sum clip(s,0,\alpha)\label{eq:loss}
\end{split}
\end{equation}
where $S_m$ is the smoothing operator which Gaussian smooths with kernel size $m$ the output flux field $s$, $d$ is the input \ZL{transmitted flux on each skewer from observation}, $R$ is a skewer-selector function which maps the 3-D field to the observed skewers, $N$ is the noise covariance matrix, $clip$ is the function clipping values outside the interval to the interval edges \ZL{and $\sum$ sums over the pixels of the clipped field}. These clipping functions have the effect of penalizing extremal and/or non-physical values, and $\alpha$ is an empirical constant between 0 and 1 depending on the mean flux and smoothing scale (which depends on spectra resolution and sightline spacing). \ZL{Our data $d$ is not required to be fixed to the grid (i.e. sightlines can be between pixel centers and values interpolated via the skewer-selector $R$.}

It is useful to note that the first two terms of this optimization procedure alone will reconstruct the standard Wiener Filter result for the case of $L_\parallel = L_\bot = m$; both terms have the same effect of penalizing the resulting likelihood for small-scale variations below that scale. However, it yields significant improvement over existing WF implementations in computational cost and memory by virtue of performing a local smoothing operation rather than a full non-sparse matrix inversion (for additional general discussion, see \citet{2019JCAP...10..035H}).

We have adjusted our optimization to perform a multiscale, annealed optimization as implemented in \citet{horowitz2020tardis}. At each step in the optimization, we smooth the resulting flux field in steps, progressing from 1.25 $h^{-1}$Mpc smoothing to  0.5 $h^{-1}$Mpc smoothing,
with the step of 0.25 $h^{-1}$ Mpc. The smoothing range we use was found through empirical testing to give the best reconstruction (minimal difference between the real map and reconstructed map). It should be noted that if the final smoothing scale is too small or zero, there will be many fake small structures in the reconstructed map, because, in such case, the first term of $\mathcal{L}$ is close or equivalent to zero and then the second term of $\mathcal{L}$ simply tries to make pixels on skewers equal to input data regardless of noise, and 0.5 $h^{-1}$Mpc smoothing in the last step is sufficient to avoid such problems. This anneal scheme helps us to approach the global minimum of $\mathcal{L}$, without getting the optimizer stuck in local minima. With sufficient tests, we get evident smaller mean squared error (MSE) in flux compared to that in optimization with a single smoothing scale 0.5 $h^{-1}$Mpc even though both of them reach numerical convergence.


\ZL{ORCA is written within TensorFlow using SciPy’s L-BFGS optimizer. L-BFGS is a quasi-Newtonian solver approximating the second derivative information, allowing quick convergence with limited memory.  We note that the L-BFGS optimizer has comparable time scaling with PCG used in {\tt dachshund}, but in practice, L-BFGS is 2-3 times faster than PCG to solve these type of problems with the same system architecture \citep{Seljak_2017}. We tested ORCA and {\tt dachshund} on CPU, we find ORCA takes comparable time to {\tt dachshund}, since we used more iterations in annealing steps, which makes ORCA slower than directly optimizing the loss. However, TensorFlow allows ORCA running on GPU for much faster optimization. In this paper, we use an NVIDIA Tesla V100S GPU 32 GB and an Intel(R) Xeon(R) Gold 6226R CPU @ 2.90GHz. We find ORCA runs 10-100 times faster than {\tt dachshund} owing to GPU acceleration.}

\section{Mock surveys}\label{sec:mock surveys}
\subsection{Mock datasets}\label{sec:datasets}
We use a hydrodynamical simulation with Nyx code \citep{Almgren_2013} for our mock survey in this paper, which has a 100 $h^{-1}$Mpc box size with particle resolution $4096^3$. The simulation uses a flat $\Lambda$CDM cosmology with $\Omega_m=0.3, \Omega_b = 0.047, h = 0.685, n_s = 0.965$, and $\sigma_8 = 0.8$. We downsample the simulation to particle resolution $200^3$ as the true field in our mock survey.

Following \citet{Krolewski_2018}, We randomly select skewers with a mean sightline separation $\left \langle d_{\rm{LOS}} \right \rangle=2.5h^{-1}$Mpc, comparable to CLAMATO $\left \langle d_{\rm{LOS}} \right \rangle=2.37h^{-1}$Mpc, and further, within the predicted range of the upcoming Subaru Prime Focus Spectrograph (Subaru/PFS) high redshift tomography program. \ZL{The selected skewers are then convolved with Gaussian smoothing to spectrograph resolution.} We also emulate the predicted observations of Thirty Meter Telescope (TMT) assuming $\left \langle d_{\rm{LOS}} \right \rangle=1h^{-1}$Mpc and same noise properties. We hereafter use N-PFS (PFS-like mock survey using Nyx simulation) and N-TMT (TMT-like mock survey using Nyx simulation) respectively for the two different mock surveys in this paper.

We apply the procedure provided by \citet{Horowitz_2019} and \citet{horowitz2020tardis} to add pixel noise to each skewer and model the continuum-fitting error. The noise level on each skewer is determined by drawing a S/N ratio from a distribution between a minimum and maximum S/N. From \citet{2015MNRAS.453..311S}, the distribution follows a power-law: $dn_{\rm{skewer}}/d(S/N) = S/N^{-\alpha}$, based on LBG luminosity function \citep{Reddy_2008} and observed distribution in \citet{Lee_2014b}. We use $\alpha$ = 2.7 which provides the best approximation of S/N distribution of CLAMATO and PFS.

To account for continuum misclassification error, the flux values within each skewer is offset such that the final observed flux is
\begin{equation}
   F_{\rm{obs}}=\frac{F}{1+\delta_c},
\end{equation}
with $\delta_c$ being a value drawn from a Gaussian distribution with mean 0 and width 
\begin{equation}
    \sigma=\frac{0.205}{S/N}+0.015.
\end{equation}
where the constants are fitted from CLAMATO data.

The S/N ratio for N-PFS ranges from 1.4 to 10 and that for N-TMT ranges from 2.8 to 10 following \citet{Horowitz_2019}. 

\subsection{Flux Reconstruction}\label{sec:flux recon}
We apply ORCA and Wiener Filter to the mock skewers and get the reconstructed flux fields. We use $\sigma_F^2=0.082,L_\bot=\langle d_{\rm{LOS}}\rangle=2.5h^{-1}\rm{Mpc,\ and}\ L_\parallel=2h^{-1}{\rm Mpc}$ in Equation(\ref{con:wiener}) for Wiener Filter reconstruction and $k_1=5,k_2=0.3,k_3=0.025$ in Eq. \ref{eq:loss} for the ORCA reconstruction. 
We use the same ORCA parameters for both mock survey and CLAMATO survey discussed in Section \ref{sec:clamato data}. \ZL{We discuss how we choose parameters of ORCA in Appendix \ref{sec:choice_param}}. While \citet{Lee_2018} uses $\sigma_F^2=0.05$ for WF in CLAMATO data, we adjust $\sigma_F^2$ to 0.082 in our mock survey to match the PDF of CLAMATO \dr. \ZL{Typically {\tt dachshund} takes $\sim$8000 seconds and ORCA takes $\sim$250 seconds to reconstruct the flux field in N-PFS mock survey.}
We apply Gaussian-smoothing to the output field with a Gaussian kernel of 2$h^{-1}$ Mpc in the following analysis
except for the void finding discussed in Section \ref{sec:vc finding}.

\begin{figure}[t!]
\includegraphics[width=0.465\textwidth]{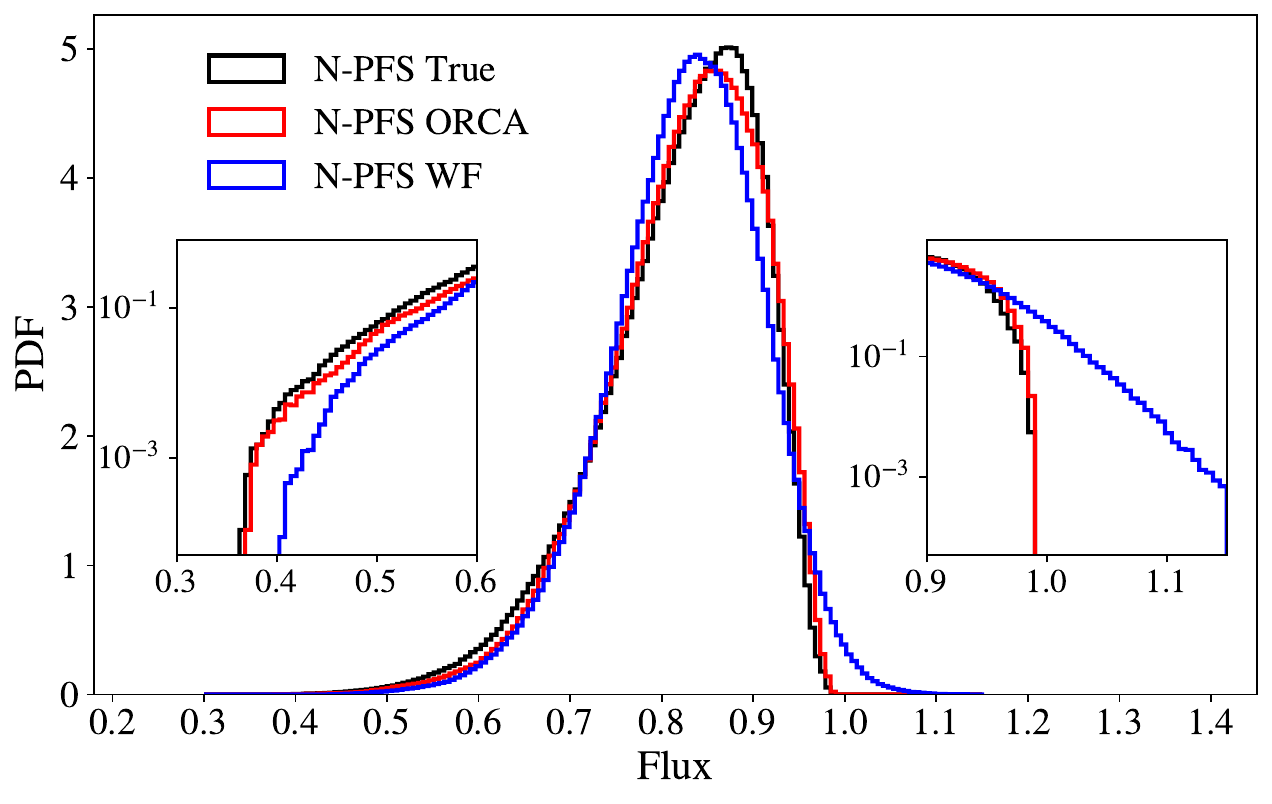}
\caption{Frequency distributions of flux in true map, Wiener filtered map and ORCA reconstructed map. The small plots inside the figure show low counts regions in the log scale. \label{fig:pdf}}
  \vspace*{\floatsep}

\includegraphics[width=0.465\textwidth]{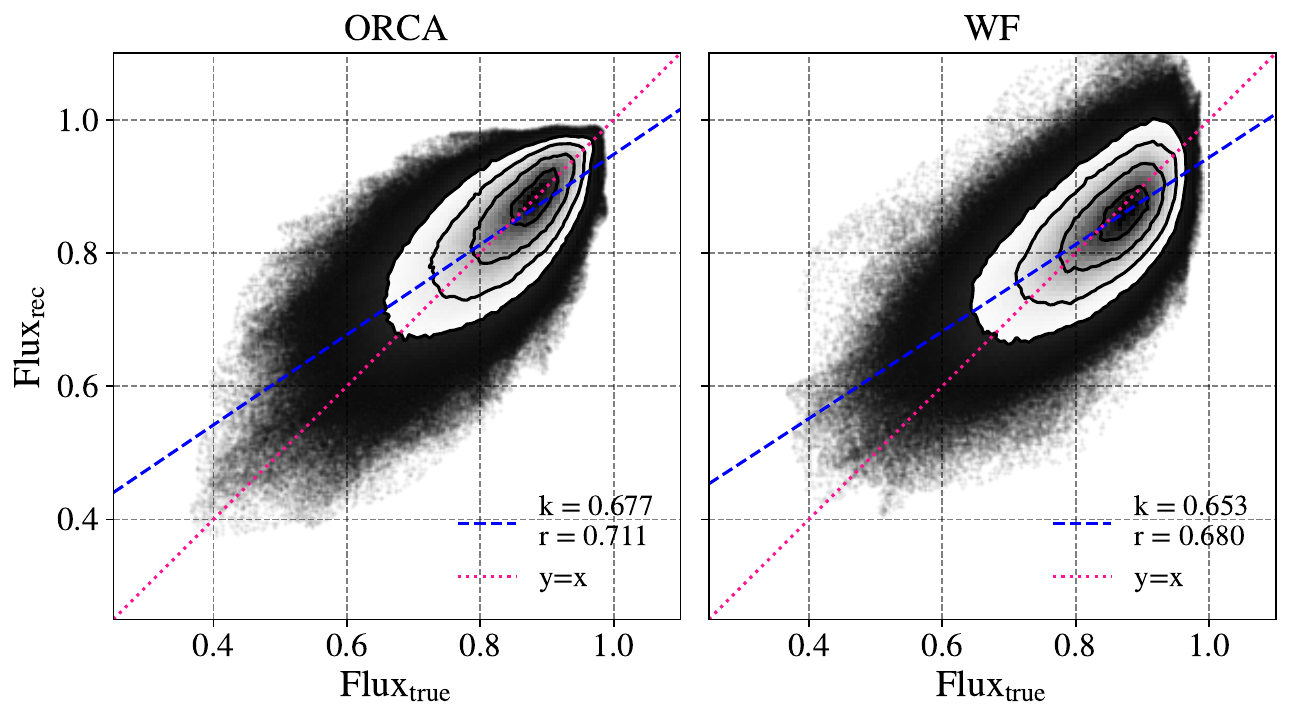}
\caption{Scatter plot of the reconstructed flux against the true flux in Nyx simulation. The red dotted line represents the Flux$_{\rm{recon}}=$ Flux$_{\rm{true}}$ relation while the blue dash line is the best linear fit of the points with $k$ slope and $r$ Pearson coefficient of the fit. These metrics show that ORCA provides a less biased, smaller variance estimate, compared to the WF solution. \label{fig:flux_scatter}}
\end{figure}

Figure \ref{fig:pdf} shows the probability density distribution of smoothed flux in the true map, Wiener filtered map and ORCA reconstructed map. The smoothed true field is quite non-Gaussian, indicating that there is more cosmological information beyond the two-point correlation function \citep{Krolewski_2018}. 
We find that the distribution of ORCA is more consistent with the true distribution and that of the Wiener filter, which
has a larger deviation. It is also notable that part of the flux values from the Wiener filter are beyond one which is non-physical 
and this is a ubiquitous pattern for Wiener filter since it has no additional constraints.
ORCA improves the flux PDF on both high and low flux tail, providing more realistic flux values in the reconstructed field.

We show scatter plots of the reconstructed flux against the true flux in Figure \ref{fig:flux_scatter}. 
We find that the relations between Flux$_{\rm{recon}}$  and Flux$_{\rm{true}}$ are biased, which is also found by \citet{Lee_2014} and \citet{8197625}. We note that they plot with flux contrast, but we use flux instead to better illustrate points with flux value beyond one. ORCA has an obvious improvement in reconstructing flux at 0.8 to 1.

After applying linear regression to the scattering points, we find a better slope and Pearson coefficient for ORCA, which are 0.677 and 0.711 compared to 0.653 and 0.680 for Wiener Filter respectively. While \citet{8197625} linearly corrected the flux according to the slope of the regression (subtract flux value by fitting y-intercept and divide them by slope), we do not correct the bias 
in the following analysis, because our cosmic web classification procedure 
(see Section \ref{sec:cosmic web}, Eq. \ref{eq:deformation_tensor}) is the second order in nature and our 
results do not change with a first order correction. 
Besides, we find that the linear correction will cause additional flux values exceeding one as the 
correction rotates the points in Figure \ref{fig:flux_scatter}. It also causes the flux distribution 
deviating from the true distribution in Figure \ref{fig:pdf}, which meets our expectation since ORCA 
has already optimized the field to get the minimal difference from the true field and 
such a simple correction would break the optimized state.


\subsection{Cosmic Web Classification}\label{sec:cosmic web}

We choose the Tidal Shear Tensor (T-web) method \citep{2009MNRAS.396.1815F} to classify the cosmic structures to test how well our method reconstructs the field.
\begin{figure}[t!]
    \includegraphics[width=0.465\textwidth]{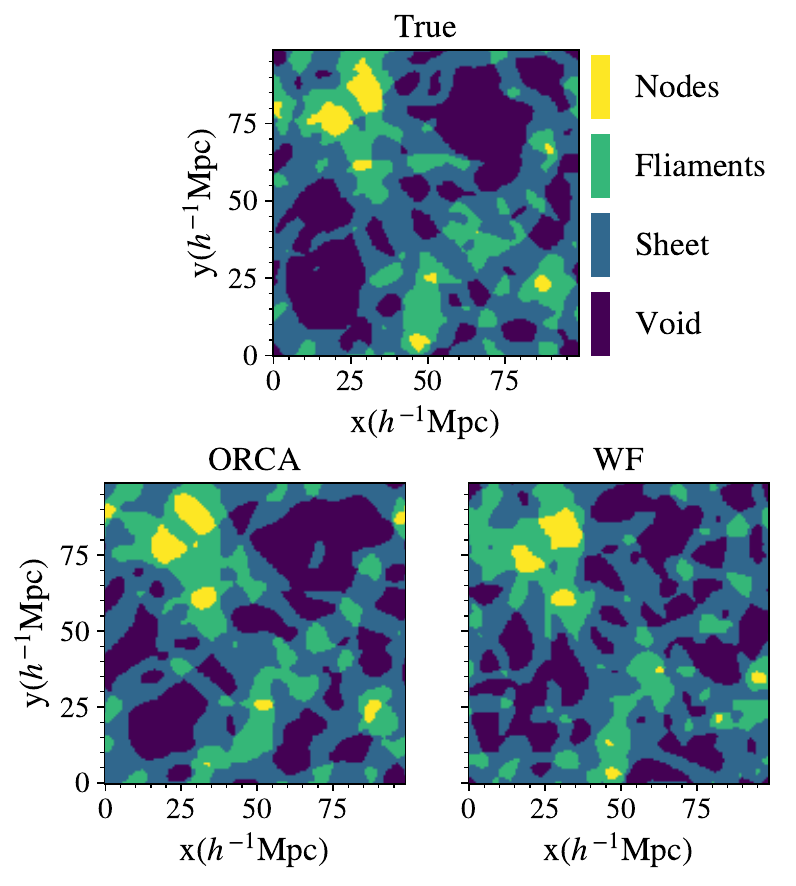}
    \caption{Cosmic Web Structures from a single slice. Void, sheet, filaments and nodes are marked in dark blue, light blue, green and yellow respectively. All maps are smoothed with a $2h^{-1}$Mpc Gaussian kernel.\label{fig:web}}
    \vspace*{\floatsep}
    \includegraphics[width=0.465\textwidth]{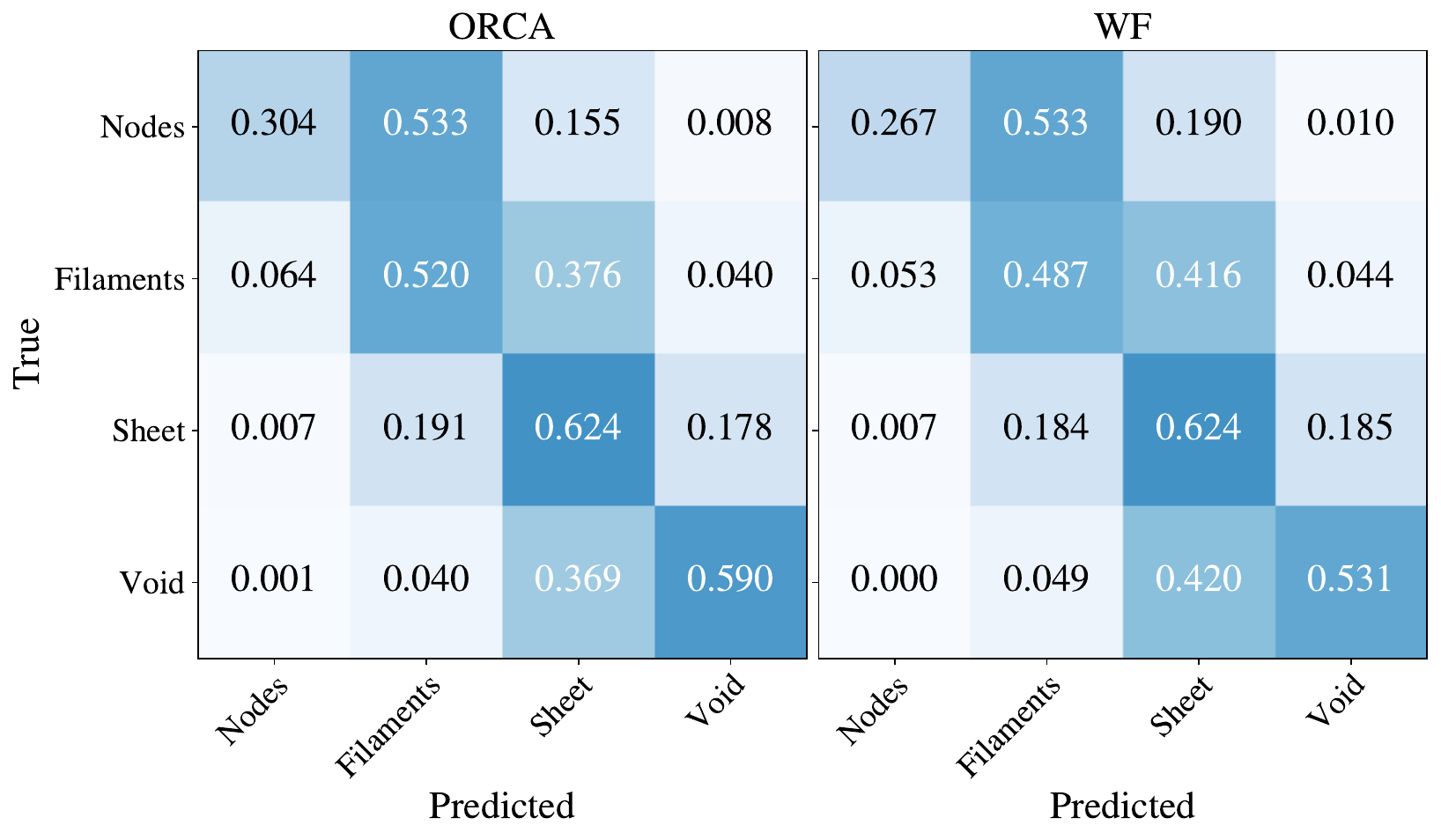}
    \caption{Confusion matrix for cosmic web classification for the N-PFS mock survey. Numbers on the diagonal the fraction of each true cosmic structure correctly identified by the reconstruction, while those off the diagonal reflect the faction of each structure misidentified. \label{fig:confusion}}
\end{figure}
The T-web method uses the deformation tensor, the Hessian of the underlying gravitational potential,
\begin{equation}
    D_{i j}=\frac{\partial^{2} \Phi}{\partial x_{i} \partial x_{j}}
    \label{eq:deformation_tensor}
\end{equation}
which is numerically practical to compute in Fourier space,
\begin{equation}
    \tilde{D}_{i j}=\frac{k_{i} k_{j}}{k^{2}} \delta_{k}
\end{equation}
where $\delta_{k}$ is the density field and we obtain $D_{i j}$ by inverse-Fourier transforming $\tilde{D}_{i j}$.

\begin{figure*}[t!]
    \plotone{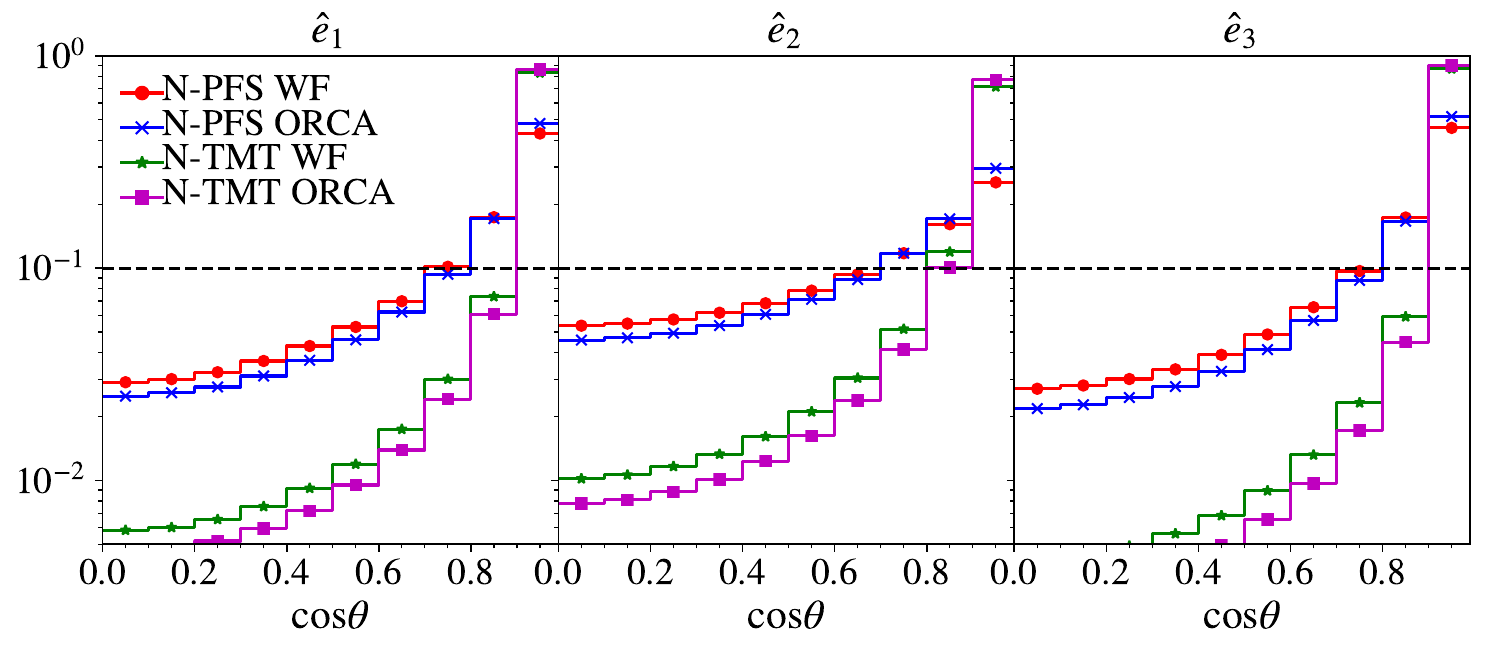}
    \caption{Histogram of the cosine of the angle between the reconstructed eigenvectors and true eigenvectors for each eigenvector. Black dashed line would correspond to random orientations of reconstructed eigenvectors, while the furthest right bin corresponds to complete agreement of eigenvectors (i.e. cos$\theta\sim1$). All the eigenvectors are computed from maps smoothed with a $2h^{-1}$Mpc Gaussian kernel.\label{fig:cosine}}
    
\end{figure*}
\begin{deluxetable*}{cccccccccc}[t]
\tabcaption{Cosmic Web Recovery\label{tab:Cosmic Web Recovery}}
\tablehead{\multirow{2}{*}{Mock Data}&\multirow{2}{*}{Method}&\multicolumn{3}{c}{Pearson Coefficients}& &\multicolumn{4}{c}{Volume overlap}\\
\cline{3-5}\cline{7-10}& &$\lambda_1$&$\lambda_2$&$\lambda_3$& & Node & Filament & Sheet & Void
}
\startdata
\multirow{2}{*}{N-PFS}& WF & 0.473 & 0.545 & 0.659 & & 0.267 & 0.487 & 0.624 & 0.531\\
&ORCA & 0.482 & 0.592 & 0.697 & & 0.304 & 0.520 & 0.624 & 0.590\\
\hline
\multirow{2}{*}{N-TMT}& WF & 0.856 & 0.902 &  0.948 & & 0.649 & 0.781 & 0.827 & 0.804\\
&ORCA & 0.882 & 0.925 & 0.962 & & 0.720 & 0.811 & 0.851 & 0.843\\
\enddata
\end{deluxetable*}
The eigenvectors of the deformation tensor relate to the principle curvature axes of the density field at each point, corresponding in the Zel’dovich approximation with the principle inflow/outflow directions. The corresponding eigenvalues determine if the net flow is inward or outward. Points with three eigenvalues above threshold value $\lambda_{th}$ are classified as nodes, two values above $\lambda_{th}$ are filaments, one value above $\lambda_{th}$ are sheets, and zero values above $\lambda_{th}$ are voids. While the density field is related to the flux field with high flux indicating low density and low flux indicating high density, we use flux field in the deformation tensor computation as describe in \citet{Lee_2016}. Thus, the relation between cosmic structures and eigenvalues above $\lambda_{th}$ is reversed (i.e. three eigenvalues above $\lambda_{th}$ are voids).

Following \citet{horowitz2020tardis}, we define our threshold value $\lambda_{th}$ for each reconstruction such that the voids occupy 22\% of the total volume. In the true flux field, we find that [22.0, 49.9, 25.4, 2.7]\% of the volume is occupied by voids, sheets, filaments, and nodes, respectively. In ORCA and WF maps of N-PFS mock survey, those numbers are [22.0, 49.2, 25.9, 3.0]\% and [22.0, 51.5, 24.1, 2.4]\%, respectively. The volume fraction of the four structures in ORCA map are closer to that in the true map compared to WF. A typical slice of the classification is shown in figure \ref{fig:web}, where we can visually find a notable improvement of ORCA for a better recovery of voids, e.g, ORCA recovers the big void at the lower-left of the slice while WF breaks it down into small voids. 

The accuracy of the cosmic web classification is further quantitatively measured by the confusion matrix in Figure \ref{fig:confusion}. The volume overlap of ORCA map with the true field for node, filament, sheet, and void, is [30.4, 52.0, 62.4, 59.0]\%, compared with [26.7, 48.7, 62.4, 53.1]\% of WF map. ORCA outperforms WF in cosmic web classification through node, filament, and void, and it's comparable to WF in sheet identification. ORCA provides the best improvement in void identification, with accuracy improved by $\approx6$\%. We sum up the total volume of the four structures correctly classified and find the volume fraction 58.0\% and 55.9\% for ORCA and WF respectively. We also include the volume overlap fraction of N-TMT survey in Table \ref{tab:Cosmic Web Recovery}. Due to finer sightline spacing and higher S/N, the classification accuracy gets fairly better for both ORCA and WF. The volume overlap is [84.3, 85.1, 81.1, 72.0]\%, compared to [80.4, 82.7, 78.1, 64.9]\% of WF, and the total volume correctly classified is 83.6\% and 80.5\% for ORCA and WF, respectively.

The alignment of the reconstructed cosmic web with true cosmic web can be studied by comparing the eigenvectors $\hat{e}_1,\hat{e}_2,\rm{and}\ \hat{e}_3$ of the pseudo-deformation tensor $D_{ij}$. Figure \ref{fig:cosine} illustrates the distribution of cos$\theta$ of the angle between the reconstructed eigenvectors and true eigenvectors. In Figure~\ref{fig:web}, we could find that ORCA has improved performance in recovering all three eigenvectors. We further increase the density of sightlines so that  $\left \langle d_{\rm{LOS}} \right \rangle=2h^{-1}$Mpc and then we find WF has a similar quality in cosmic web alignments. It indicates that reconstructing by ORCA has equivalent effects of increasing sightlines. And for N-TMT survey, ORCA still notably outperforms WF. We also quantify the agreement between reconstructed field and true field in terms of Pearson coefficients of reconstructed and true eigenvalues in Table \ref{tab:Cosmic Web Recovery}. It shows that we get a stronger correlation between three reconstructed and true eigenvalues [$\lambda_1,\lambda_2,\lambda_3$] either for N-PFS or N-TMT surveys with ORCA, which ranges from [0.482, 0.592, 0.697] to [0.882, 0.925, 0.962] in contrast with that ranging from [0.473, 0.545, 0.659] to [0.856, 0.902, 0.948] with WF.
\section{Application to CLAMATO Data} \label{sec:clamato data}

We apply our technique to the first data release of the COSMOS Lyman Alpha Mapping And Tomographic Observations (CLAMATO) survey\footnote{The data is available from: \url{https://doi.org/10.5281/zenodo.1292459}. We use map\_2017\_v4.bin and map\_2017\_v4\_sm2.0.bin as Wiener Filter map and pixel\_data\_v4.bin as input to ORCA.} \citep{Lee_2018}. This data includes reduced Ly$\alpha$ forest signatures from 240 galaxies and quasars with redshifts ranging $2.17<z<3.00$, allowing reconstruction at $2.05<z<2.55$ over $\sim0.157$ square degree. Standard Wiener filter reconstructions of this data have been used to detect a large number of cosmic voids and proto-cluster regions.

\subsection{Void and cluster finding} \label{sec:vc finding}
\begin{figure*}[t]
    \plotone{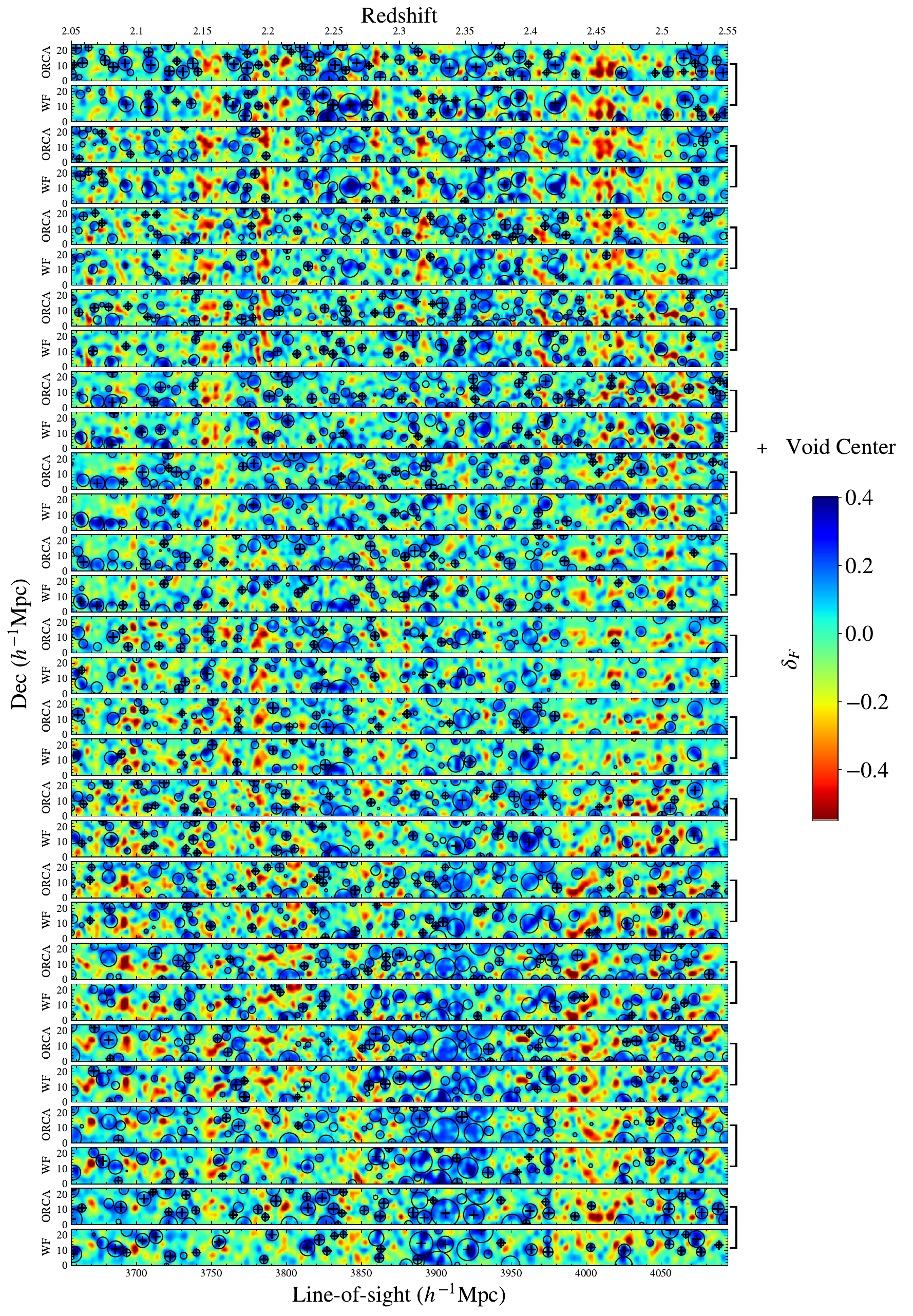}
    \caption{Comparison of voids found in CLAMATO map reconstructed by Wiener Filter and ORCA. Each strip is the stack of 4 slices with 2 $h^{-1}$Mpc in thickness through the RA direction. The black circles represent the voids intersecting the slices and the + marks all voids centered at one of four slices of a strip. The strips connected by staples on the right are reconstructions with ORCA and Wiener Filter at the same RA marked by ORCA and WF on the left. \label{fig:void compaison}}
\end{figure*}
We adopted the void finding procedure presented in \citet{2015MNRAS.453.4311S} and compared void catalogs in the map reconstructed by Wiener Filter and that by ORCA. 
In the flux contrast field, we begin by finding all points with $\delta_F$ above a threshold (SO threshold). Spheres then grow centered on all those points until the average $\delta_F$ inside the sphere reaching a second threshold (SO average). 
Due to the pixel noise and continuum error, the PDF of WF \dr is broadened, especially in high flux region where we find voids, and we use different thresholds for true flux contrast field $\delta_F$ and WF \dr. For both true and WF field in mock survey, we use those thresholds provided by \citet{Krolewski_2018}. The process of choosing thresholds can be summarized as follows (also see Table~2): 
 \begin{enumerate}[leftmargin=20pt]
   \item Find voids in the true redshift-space density field of the Nyx simulation. \label{density}
   \begin{itemize}[leftmargin=4pt]
       \item SO thresh $=0.15\Bar{\rho}$ and SO avg $=0.3\Bar{\rho}$.
   \end{itemize}
   \item Find voids in the true flux contrast field of the Nyx simulation.\label{true}
    \begin{itemize}[leftmargin=4pt]
        \item Choose SO thresh and SO avg to best match voids found in \ref{density}.
        \item SO thresh $=0.192$, SO avg $=0.152$.
    \end{itemize}
    \item Find voids in the reconstructed field.
    \begin{enumerate}[leftmargin=12pt]
        \item Wiener Filter
        \begin{itemize}[leftmargin=4pt]
            \item Choose SO thresh and SO avg to best match voids found in mock survey and in \ref{density}.
            \item SO thresh $=0.220$, SO avg$= 0.175$ for both mock survey and CLAMATO.
        \end{itemize}
        \item ORCA
        \begin{itemize}[leftmargin=4pt]
            \item Use the same thresholds in \ref{true}.
            \item SO thresh $=0.192$, SO avg $=0.152$ for both mock survey and CLAMATO.
        \end{itemize}
    \end{enumerate}
 \end{enumerate}

In the true field, the voids are firstly identified in redshift-space density field using SO threshold$=0.15\Bar{\rho}$ and SO average$=0.3\Bar{\rho}$. The thresholds for true flux contrast field $\delta_F$ and WF reconstructed field \dr are chosen to be [0.192,0.152] and [0.220,0.175] respectively, matching the void fraction in redshift-space density field. 
Nevertheless, we did not use the same threshold for ORCA \dr as that of WF, because ORCA has constraints on absorption and we find ORCA working well around the thresholds of the true field. Thus, we simply use the same thresholds for ORCA \dr.

\begin{figure}[t!]
    \includegraphics[width=0.44\textwidth]{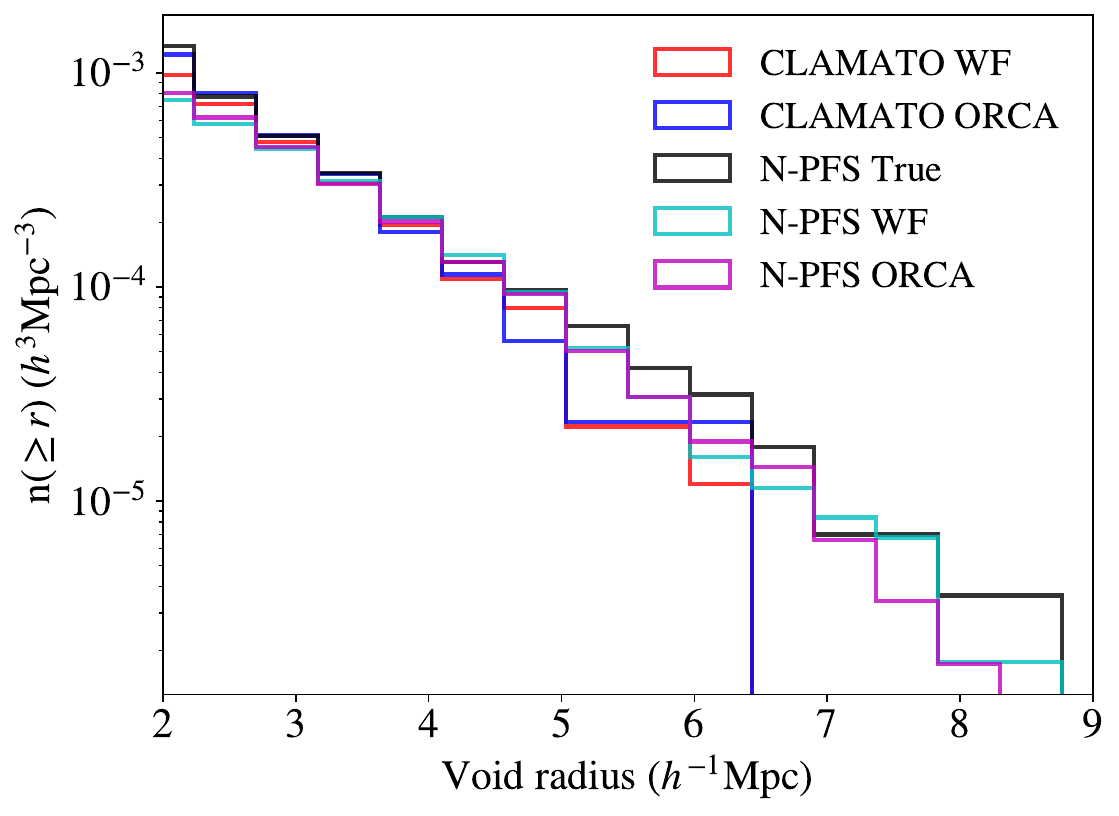}
    \caption{Void radius function for our mock catalog (NyX) and the CLAMATO data. Due to the geometry of the CLAMATO volume, there are few large voids ($r > 6.5$ $h^{-1}$ Mpc) identified.\label{fig:radius}}
    \end{figure}

 Qualitatively, we find a good agreement between the two CLAMATO maps\footnote{For the Wiener Filter map, we use the void catalog from: \url{https://doi.org/10.5281/zenodo.1295839}.}. While \citet{Krolewski_2018} identified 355 $r> 2 h^{-1}$ Mpc voids, including 48 higher-quality $r\ge5$ voids, we find 496 $r>2 h^{-1}$ Mpc voids and 55 $r\ge5\ h^{-1}$ Mpc voids. We find that there is 70.52\% of the volume of voids found in the Wiener Filtered map that is also found in ORCA reconstructed map. This high overlap fraction indicates a good agreement between the two methods. We plot the comparison of voids found in two maps in Figure \ref{fig:void compaison}. The figure shows the stack of every four slices with 2 $h^{-1}$ Mpc in the RA or DEC direction. We also compare the void radius function in CLAMATO to that in the N-PFS mock survey of the two methods in Figure \ref{fig:radius}. Due to edge effects, voids are more likely to be found near survey boundaries, so we exclude all the voids with a distance from the center to the boundary smaller than the void radius. Following \citet{Krolewski_2018}, we compute the void radius function with weights to each void by the effective volume over which it could have been observed (e.g., for the geometry of CLAMATO, the effective volume is $(30-2r)(24-2r)438$ ${\rm Mpc^3}h^{-3}$ with $r$ the void radius). We find a good agreement of void radius function in CLAMATO and mock survey, either with ORCA or WF. 

\begin{deluxetable*}{cccccc}[t]
    \tablecaption{Volume fraction for different thresholds of void and cluster in simulated and CLAMATO catalogs.\label{tab:void th}}
    \tablewidth{0pt}
    \tablehead{
    \colhead{Type} & \colhead{Data}&\colhead{Field} & \colhead{SO thresh} & \colhead{SO avg} & \colhead{Vol. frac.}
    }
    \startdata
    \multirow{5}{*}{Void}&\multirow{3}{*}{N-PFS}&$\delta_F$& 0.192 & 0.152 & 22.89\%  \\
    & &ORCA \dr& 0.192 & 0.152 & 18.37\% \\
    & &WF \dr& 0.220 & 0.175 & 17.53\% \\
    \cline{2-6}
    &\multirow{2}{*}{CLAMATO}&ORCA\dr  & 0.192 & 0.152 & 24.34\% \\
    & &WF \dr & 0.220 & 0.175 & 19.47\% \\
    \hline
    \multirow{5}{*}{Cluster}&\multirow{3}{*}{N-PFS}&$\delta_F$& -0.304 & -0.271 & 2.71\%  \\
    & &ORCA \dr& -0.304 & -0.271 & 2.11\% \\
    & &WF \dr& -0.304 & -0.271 & 1.72\% \\
    \cline{2-6}
    &\multirow{2}{*}{CLAMATO}&ORCA \dr & -0.304 & -0.271 & 1.96\% \\
    & &WF \dr& -0.304 & -0.271 & 1.20\% \\
    \enddata
    \tablecomments{Comparison of volume fraction of voids and clusters found in Nyx mock survey (100$h^{-1}$Mpc box) and CLAMATO survey with two methods. All the maps for finding clusters are smoothed with a 2 $h^{-1}\rm{Mpc}$ Gaussian kernel, while that for finding voids are without additional smoothing.}
\end{deluxetable*}

To test the improvement of ORCA in void recovery, 
we define the void volume overlap completeness as 
the fraction of voids found in the true flux map from our mock catalog that 
are also found in reconstructed map, and the volume 
overlap purity as the fraction of voids found in 
reconstructed map that are also found in the true flux map, i.e. 
\begin{align}
    \textrm{Purity} \equiv & \frac{V_{\rm{true}}\cap V_{\rm{rec}}}{V_{\rm{rec}}}, \label{eq:purity}\\
    \textrm{Completeness} \equiv & \frac{V_{\rm{true}}\cap V_{\rm{rec}}}{V_{\rm{true}}}.\label{eq:complete}
\end{align}
\ZL{Here $\cap$ denotes the volume overlap of voids between true and reconstructed catalogs. We only consider the volume overlap, regardless of shifts of center position and radius (i.e., if voids in two catalogs are different in size and center position but they share overlapped volume, we calculate such volume for purity and completeness).}
In the N-PFS mock survey, we find that ORCA reconstructed map has a higher volume overlap completeness computed using all voids, which is 34.50\% compared to 30.72\% in the WF reconstructed map.
\begin{figure}[t!]
    \plotone{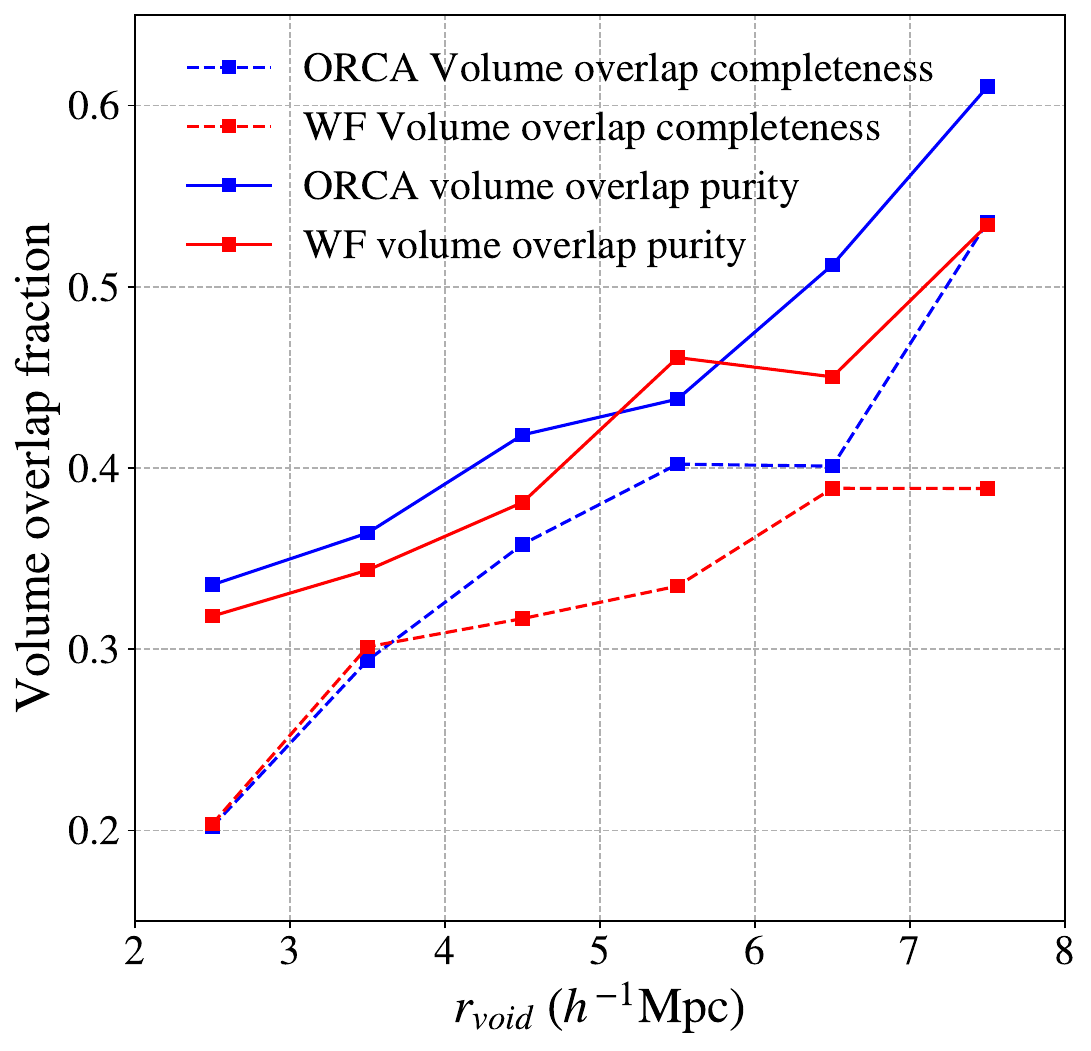}
    \caption{Purity (solid line) and completeness (dashed line) of volume overlap fraction of N-PFS mock survey, as defined in Eq, \ref{eq:purity} and Eq. \ref{eq:complete}. Each bin is 1$h^{-1}$Mpc.\label{fig:overlap}}
    \end{figure}
\begin{figure*}[ht!]
    \plotone{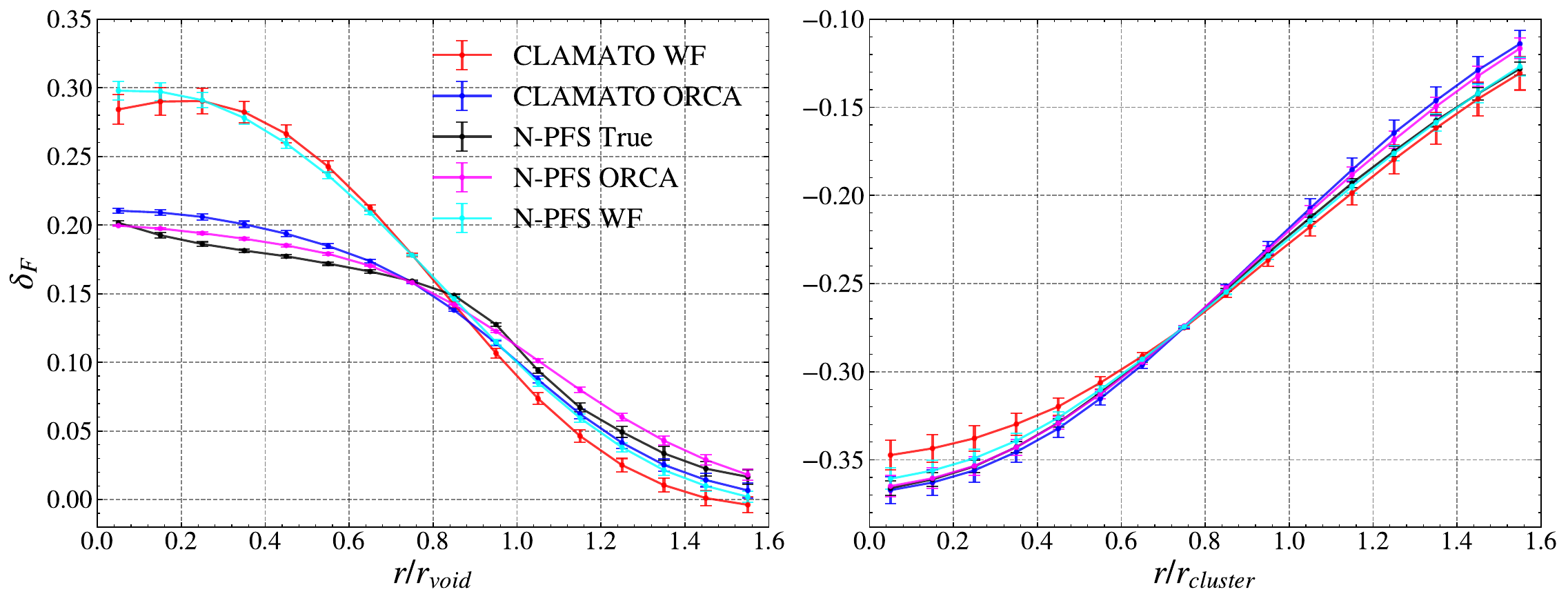}
    +\caption{Solid lines represent the mean of void (left) and cluster (right) profiles in  Wiener reconstructed map from CLAMATO data (red), ORCA reconstructed map from CLAMATO data (blue), Nyx map (black), ORCA reconstructed map from Nyx mock observations (magenta) and Wiener reconstructed map from Nyx mock observations (cyan) and errorbars show standard deviation $\sigma$ of the stacks. \label{fig:vc profile}}
\end{figure*}
In Figure \ref{fig:overlap}, we plot the completeness and purity of the volume overlap fraction compared between voids in mock survey reconstructed by ORCA and WF, and the true flux field in the Nyx simulation as a function of void radius. While for small voids ($r<4$), the completeness is comparable for both Wiener Filter and ORCA, but one can see a substantial improvement for ORCA as void radius increases. ORCA also generally outperforms the Wiener Filter for volume overlap purity especially for large voids ($r>6$). We notice that ORCA identified more voids together with larger void volume fraction in both mock survey and CLAMATO data, as can be seen in Table~\ref{tab:void th}, and with the improvement of both purity and completeness verified in the mock survey, the void catalog of CLAMATO map recovered with ORCA should be more authentic.

We plot the radically-averaged void profiles at the left panel of Figure \ref{fig:vc profile} for all voids with $r \ge 5 h^{-1}$Mpc, normalizing each void to its void radius and stacking in units of the void radius $r/r_{void}$. We could see a good agreement between void profiles in the mock survey and CLAMATO data. The void profiles for $r>r_{void}$ in mock surveys trace well the true flux contrast field $\delta_F$. \ZL{However, we find a large deviation between void profiles in WF \dr\ and Nyx $\delta_F$, while void profile in ORCA \dr\ matches better with the profile in Nyx $\delta_F$. ORCA provides a better reconstruction of voids, and we can study the environments inside voids more accurately.}

We also apply SO method to find clusters using a similar procedure. We smooth both true and reconstructed maps with a 2 $h^{-1}$Mpc Gaussian kernel following \citep{2015MNRAS.453..311S}. The volume fraction of nodes in the true nyx map classified by T-web method (see Section \ref{sec:cosmic web}) is 2.7\% and we choose thresholds for SO cluster finder to make the volume fraction of clusters in the nyx map match this fraction. We use the same thresholds for both true and reconstructed maps. \ZL{Similarly, we only take into account clusters with $r \ge 2.5 h^{-1}$Mpc for making the profiles, omitting small clusters which are more likely to be contaminated by noise}. It can be seen at the right panel of Figure \ref{fig:vc profile} that ORCA tends to recover overdensities better, while Wiener Filter underestimates the density inside the clusters.

\section{Conclusion}\label{sec:conclusion}
In this work, we have introduced a new tomographic flux reconstruction technique to use on \lya\ Forest observations. Testing our approach on mock catalogs from hydrodynamical simulations, we have shown improved cosmic web reconstruction vs. standard Wiener filtering approaches. This improvement can be seen both in classification accuracy as well as reconstruction of profiles and number statistics of voids and (proto)clusters. In addition to testing on mock catalogs, we have also applied our technique to data from the CLAMATO Survey and have found good agreement with void profiles from simulations. In our simulations, we have also found that our method can reconstruct void profiles more consistent with true void profile, providing a way to more accurately study physics inside voids from observations.

As \lya\ Tomography is expected to play a major role in upcoming spectroscopic surveys, such as Subaru Prime Focus Spectrograph (PFS) \citep{2014PASJ...66R...1T}, it is important to gain the maximum information from the limited time available. We have found that ORCA provides large scale structure constructions comparable to WF with 30-40\% more sightlines, depending on metric used. With the increasing of survey volume and sightline density in upcoming surveys, the computational costs of the WF reconstructions become more apparent. \ZL{We find that ORCA, with GPU acceleration, reconstructs 10-100 times faster than Wiener Filter ({\tt dachshund}), depending on specific surveys.}

Going forward, this technique provides a complementary tool with forward model density reconstruct techniques (e.g. \cite{horowitz2020tardis}), which rely on strong assumptions about IGM physics, and provides a useful tool to cross-correlate with galaxy properties. A regime that is of particular interest for galaxy evolution studies is the center of proto-cluster regions, where one expects significant deviations from FGPA. We hope to further explore ORCA and similar extensions to WF in this regime in future works.

\acknowledgments
 ZL and ZC are supported by the National Key R\&D Program of China (grant No. 2018YFA0404503). BH is supported by the AI Accelerator program of the Schmidt Futures Foundation. We thank Khee-Gan Lee and Alex Krolewski for helpful discussions. We thank Benjamin Zhang for help on technical issues relating to {\tt Dachshund} implementation.

This research used resources of the National Energy Research Scientific Computing Center, a DOE Office of Science User Facility supported by the Office of Science of the U.S. Department of Energy under Contract No. DEC02-05CH11231. We also acknowledge the computing resources from the Department of Astronomy at Tsinghua University.

\appendix

\section{Error Analysis on ORCA Reconstruction}
\begin{figure}[h]
\includegraphics[width=0.65\textwidth]{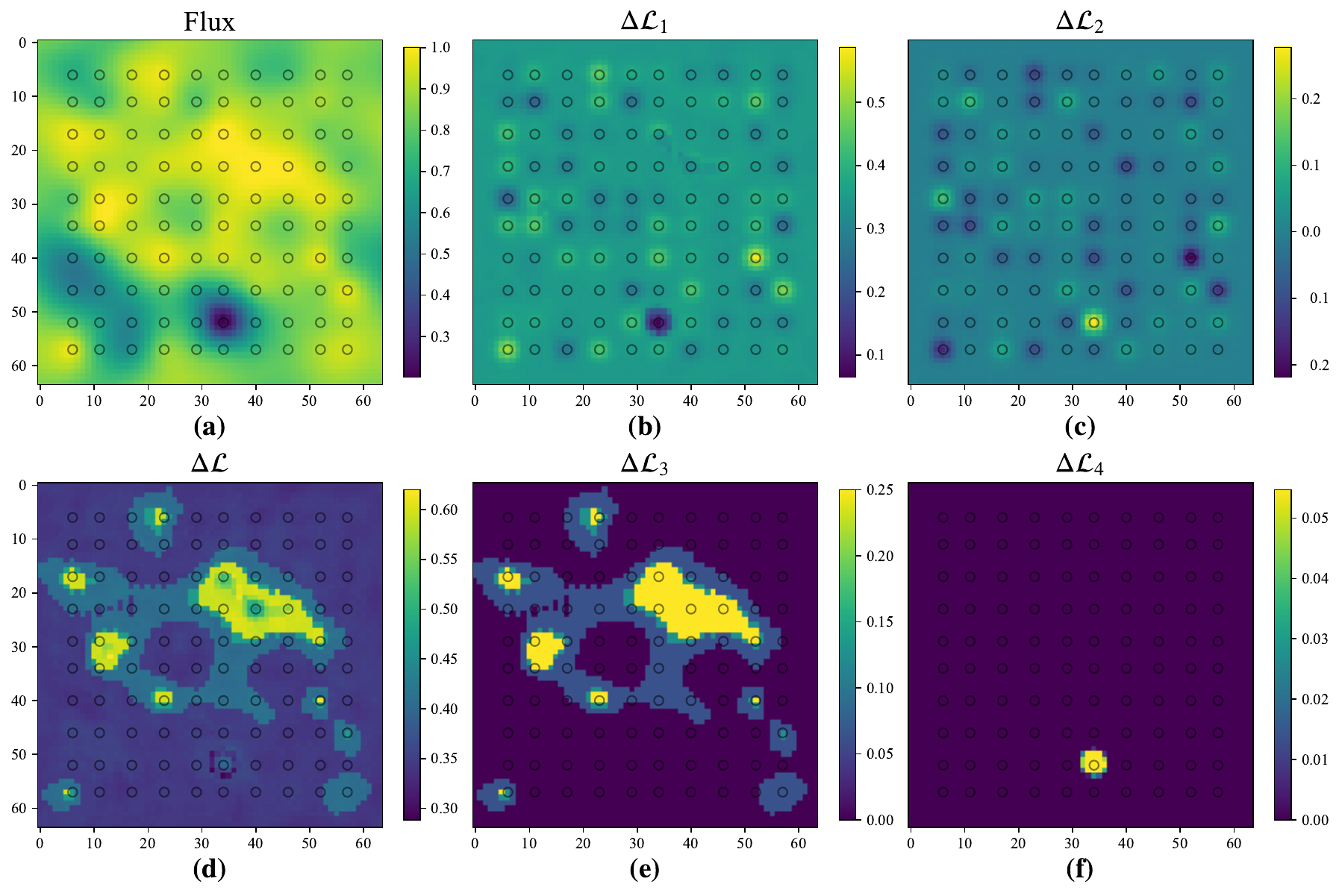}\centering
\caption{{\bf(a)} shows the original flux field from the ORCA reconstruction. {\bf (d)} shows the change of total loss function $\Delta \mathcal{L}$ in response to an increment to the flux, while {\bf(b)}, {\bf(c)}, {\bf (e)} and {\bf(f)} show the change of each separated term. Black circles indicate position of skewers.
\label{fig:error}}
\end{figure}

In order to propagate uncertainties to whatever analysis the tomographic map will be used for we need to understand the error properties of our map. Since we have a complete likelihood, it is easy to test the relative likelihood of a given flux realization. The errors will be correlated, as the signal covariance term forces the reconstructed map to be smooth. If we are interested in a pixel by pixel flux error we can calculate this via a response formalism; i.e. by varying each pixel we can study the change in the resulting likelihood. In order to know how the loss function responds, we separate the $\mathcal{L}$ in Eq. \ref{eq:loss} into 4 terms ($\mathcal{L}_1,\mathcal{L}_2,\mathcal{L}_3,\mathcal{L}_4$):
\begin{equation}
\begin{split}
\mathcal{L}_1&= k_1(S_m(s) - s)^2\\
\mathcal{L}_2&=(R(s) - d)^\mathrm{T}N^{-1} (R(s)-d)\\
\mathcal{L}_3&=k_2\sum clip(s,1,+\infty)\\
\mathcal{L}_4&=k_3\sum clip(s,0,\alpha)\\
\end{split}
\end{equation}
We add a small increment to the flux at every pixel and see how the value of each term change to penalize the optimization. Figure \ref{fig:error} shows the flux field and the change of loss function $\mathcal{L}$ and its four components at a slice perpendicular to the LOSs, added 0.1 to the flux at each pixel. We can see in {\bf(b)} and {\bf(c)} that the pixels on skewers are the most important ones with the biggest impact on the first two terms compared to other pixels off the skewers, which is expected in standard Wiener Filter as it uses information around skewers. We see the effect of ORCA in {\bf(e)} and {\bf(f)} where we add two constraints to the optimization. In {\bf(e)} we see that $\mathcal{L}_3$ works when flux values exceeding one and the yellow and light blue regions can correspond to high flux region in {\bf(a)}. With such a constraint, we could avoid the non-physical values in our final optimized map. In {\bf(f)} we see that $\mathcal{L}_4$ works at low flux region, and it penalizes the optimization when we lose some low flux values, which helps us better recover overdensity. It can also prove that the optimization has reached the minimum of $\mathcal{L}$ where any increment to flux will increase the total loss function, shown in {\bf(d)} where all $\Delta\mathcal{L}$ are positive.

\section{\ZL{The choice of parameters in loss function}\label{sec:choice_param}}
\begin{figure*}[h]
\plotone{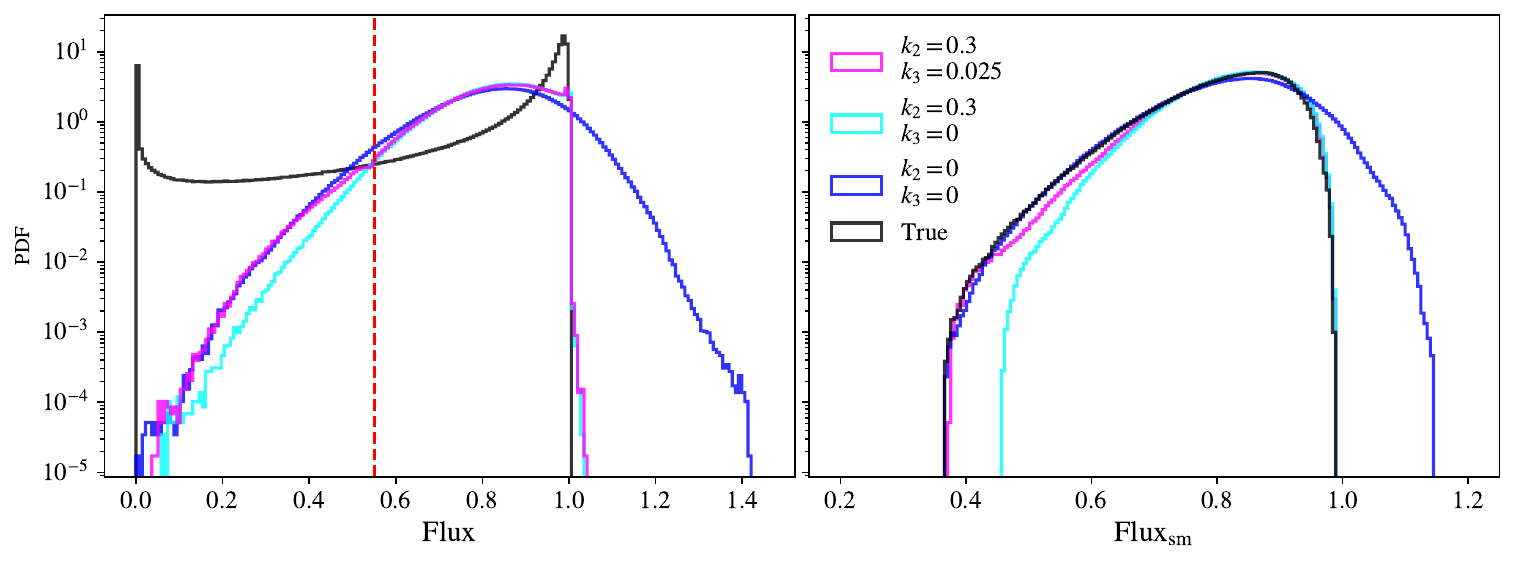}
\caption{\textit{left panel}: The probability distribution function of true and reconstructed flux in Nyx using different loss functions in ORCA. The red dashed line denotes our choice of $\alpha$ located at $\rm Flux=0.55$. \textit{right panel}: the same as the left but with additional $\sigma=$2$h^{-1}$Mpc Gaussian smoothing for all the fields. \label{fig:loss_func_comp}}
\end{figure*}

\ZL{To explain how we choose parameters in ORCA, we run the mock survey again with the same properties discussed above. $k_1$ works as adding smoothing to the output field, and it is chosen empirically to make the field looks less noisy. We find $k_1=5$ works best for the S/N and mean sightline spacing in our problem. 
We use the term with $k_2$ aiming at penalizing the optimization for transmitted flux values above one.  Nevertheless, it also impacts the optimization for somehow reducing the low flux values which correspond to overdensity. To compensate for overdensity in the map, we add terms with $k_3$ and $\alpha$. We keep $k_1=5$, varying $k_2$ and $k_3$ to illustrate their contributions to the field. At the left panel of Figure \ref{fig:loss_func_comp}, we could see that the ORCA optimized transmitted flux field with $k_2=0.3,k3=0$ (cyan) underestimates the overdensity below $\rm Flux\approx0.55$, compared to the true transmission field from Nyx simulations (black). While with $k_2=0.3,k3=0.025$, the ORCA optimized field gets compensated for overdensities, which can be seen on both left and right panel of Figure \ref{fig:loss_func_comp}. The choice of $k_2$ and $k_3$ is also empirical, and they should not weigh too high in the loss function as the first two terms are more important in reconstructing the field. We find it useful to set $k_2$  an order magnitude smaller than $k_1$, and $k_3$ an order magnitude smaller than $k_2$. We choose $\alpha$ to a transmitted flux value where we start to underestimate the transmission. We finally use the parameters which make the reconstructed flux PDF matches best with the true flux in Nyx simulation, and apply them to both mock surveys and real observations. One could expect to alter those parameters depending on the specific problems considered. E.g., the true transmitted flux evolves with redshift changing the PDF of true flux, and we should test where we may underestimate the flux as to pick up a different $\alpha$ at different redshifts. Also, when using a different resolution, sightline spacing or $k_1$, the reconstructed flux PDF could be altered, and a different $\alpha$ is needed to decide where to compensate.}

\section{\ZL{robustness}}
\begin{figure}[t]
\plotone{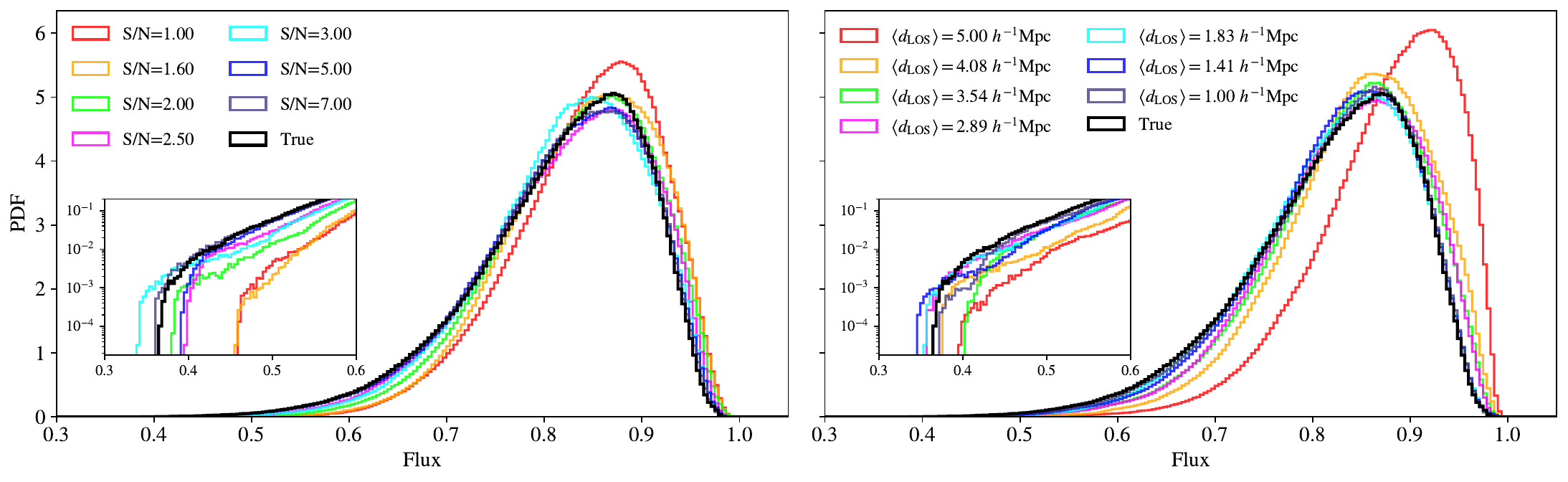}
\caption{The PDF of ORCA reconstructed flux with different S/N (left) and mean sightline spacing (right).
\label{fig:robustness}}
\end{figure}

\ZL{To test our algorithm's robustness, we run the ORCA reconstruction varying the mean sightline spacing and S/N using fixed parameters $k_1,k_2,k_3$ and $\alpha$. We first vary S/N from 1 to 9  (the same S/N for all skewers in one mock) with a constant $\langle d_{\rm LOS}\rangle=2.5h^{-1}$Mpc. We then vary $\langle d_{\rm LOS}\rangle$ from 1 to $5h^{-1}$Mpc with the same S/N distribution as in N-PFS mock. At the left panel of Figure \ref{fig:robustness}, we find that the quality of reconstructed flux PDF matches true PDF increasingly better as we increase the S/N. ORCA performs well when $\rm S/N>2$, while we still get reasonably good results with $\rm 1<S/N<2$. At the right panel of Figure \ref{fig:robustness}, we find ORCA works well with all $\langle d_{\rm LOS}\rangle$ tested except for $\langle d_{\rm LOS}\rangle=5h^{-1}$Mpc, whose sightlines are too sparsely sampled.}

\ZL{
We find ORCA is robust with parameters we choose in this paper, as the flux PDF will not be influenced much and we still get reasonable map quality when using different sightline spacing and S/N at fixed parameters. We only need to adjust those parameters around the ones we give in this paper for different problems to get optimal results, without searching for parameters in a wide range.}

\bibliography{sample63}{}
\bibliographystyle{aasjournal}

\end{document}